\documentstyle[aps,psfig]{revtex}

\newcommand{\be}{\begin{equation}}
\newcommand{\ee}{\end{equation}}
\newcommand{\bex}{\begin{eqnarray}}
\newcommand{\eex}{\end{eqnarray}}
\begin{document}

\begin{titlepage}
\title{Do quantum nonlocal correlations imply information transfer?- \\
A simple quantum optical test.}
\author{R. Srikanth\thanks{e-mail: srik@iiap.ernet.in} \\
Indian Institute of Astrophysics, Koramangala, \\
Bangalore- 34, Karnataka, India.}
\maketitle
\date{}

\pacs{03.65.Bz,03.30.+p}

\begin{abstract}
In order to understand whether nonlocality implies information transfer, a 
quantum optical experimental test, well within the scope of current technology,
is proposed. It is essentially a delayed choice experiment as applied
to entangled particles. The basic idea is: given two observers sharing 
position-momentum entangled photons, one party chooses whether she measures 
position or momentum of her photons after the particles leave the source. The 
other party should infer her action by checking for the absence or 
presence of characteristic interference patterns after subjecting his particles 
to certain optical pre-processing. 
An occurance of signal transmission is attributed to the breakdown of
complementarity in incomplete measurements. Since the result implies that 
the transferred information is classical, we discuss some propositions
for safeguarding causality.
\end{abstract}
\end{titlepage}
 
\section{Introduction}

Quantum information has opened up a new era in recent times both in
fundamental and applied physics. Its nonclassical resources of quantum
superposition and entanglement are at the heart of powerful future applications
in communication \cite{ben93} and computation \cite{preskill}.
And yet, the fundamentally very important question whether the correlated
measurements on entangled systems imply an "action-at-a-distance" effect 
remains somehow unclear. Einstein, Podolsky and Rosen (EPR) thought that it
did, which was the basis of their claim of quantum mechanical incompleteness
\cite{epr}.
Quantum nonlocal correlations have been confirmed in
experiments since the mid-1980's performed both on spin entangled systems
\cite{bellexp}, based on the Bohm version \cite{bohm} of the EPR 
thought-experiment, which are shown to violate Bell's inequality
\cite{bella}, and also on systems entangled in continuous variables
(Refs. \cite{ghosh,str95} and references therein), 
where nonlocality is manifested in multi-particle interferences. 

Bell's celebrated theorem \cite{bella} tells us
only that any realistic model of quantum mechanics should be nonlocal.
Informed opinions diverge between on the one hand the view that quantum
nonlocality implies
no information transfer, but only a change in the mutual knowledge of the two
nonlocal systems, to the acknowledgement on the other hand of a tension
between quantum theory and special relativity \cite{weihs}. The tension stems
from the possibility that the nonlocal correlations might imply a 
superluminal transfer of information. A majority of physicists in the field, 
it would seem, accept the scenario of a spacelike but causal enforcement of
correlation, as for example in quantum dense coding \cite{dense}.
In this view, entanglement cannot be used to transmit classical signals
nonlocally  
because statistically the single particle outcome at any one particle is not 
affected by measurements on its entangled parties
\cite{nosig}. This understanding is echoed in statements of 
``a deep mystery" \cite{ghz}, and ``peaceful coexistence" \cite{shi89}
between quantum nonlocality and special relativity. 

In the present article, we propose a simple quantum optical experiment
whose aim is to test in a philosophically unpredisposed 
way whether information transfer occurs in nonlocal systems. 

\section{A practical experiment}\label{action}

\begin{figure}
\centerline{\psfig{file=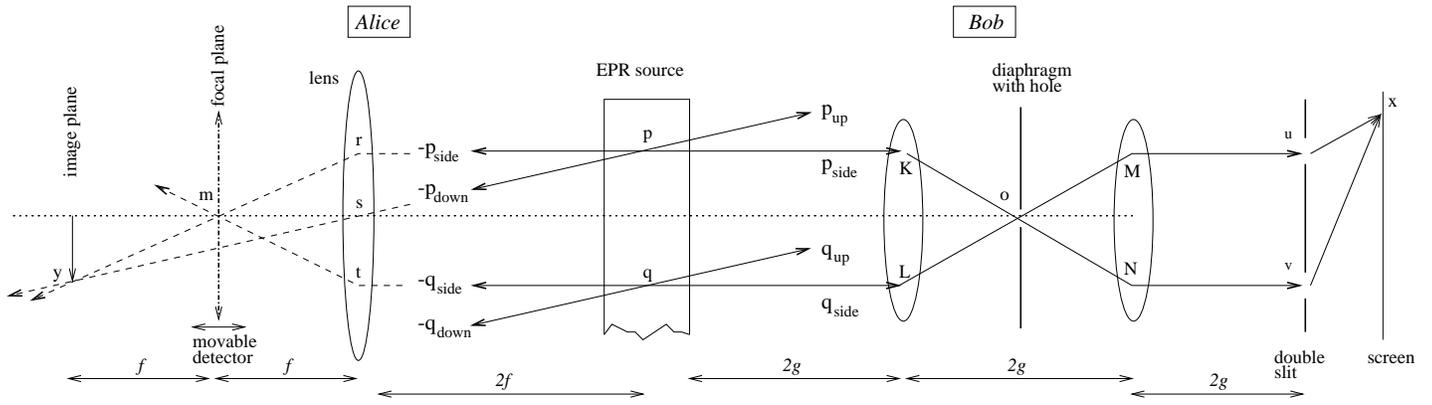}}
\caption{Light
produced in a nonlinear crystal pumped via spontaneous parametric 
downconversion (SPDC) is shared by Alice and Bob. Alice measures the
position or momentum of her photons by detecting them at the image- or 
focal-plane of her
lens. The interference pattern produced by their twins in the single counts is 
observed by Bob using a double-slit interferometer. Access to Bob's
interferometer is restricted to horizontal rays, by means of two lenses
facing each other with a single-hole perforated diaphragm placed at their 
common focal plane, so that only horizontal rays fall on the interferometer. 
The intensity and visibility of his 
interference pattern can be controlled by Alice's choice of observation. }
\label{rayent}
\end{figure}

Figure \ref{rayent}
presents a `folded out' plan of an experiment in which two observers, 
designated Alice and Bob, share
entangled photons from a nonlinear crystal pumped by 
a suitable laser (eg., Ar laser in $\lambda$ 351.1 nm). The correlated photons 
are produced by spontaneous parametric down-conversion (SPDC) \cite{str95}. 
Photons not down-converted are filtered out (not shown in Figure \ref{rayent}),
leaving only entangled photon pairs to be shared between Alice and Bob.

Alice's observes her photons through a lens of focal length $f$. 
It is positioned at distance $2f$ from the
EPR source. By classical optics, coplanar rays that are parallel in front of 
her lens converge to a single point on the focal plane. 
A detection by Alice at some point on the focal plane on her side of the lens
implies that Bob's photon is left in a definite momentum state, 
but with its point of origin in the source indeterminate as expected
on basis of the uncertainty principle. This has indeed been 
confirmed by observing an interference pattern in Bob's photons detected in 
coincidence with Alice's momentum measurement \cite{zei00}.

On the other hand, by positioning her detector at some point on the
image plane of her lens, Alice images the source, and hence measures the point 
of origin of her photon, but here the momentum with which it left the source 
remains indeterminate. Bob's photon is also left in a position eigenstate, with
the position of its detection being correlated with that of Alice's 
coincidentally detected photon \cite{zei00}. Bob is equipped with
a Young's double-slit interferometer and a direction filter permitting only
horizontal momenta to reach Bob's interferometer. The filter consists of two
convex lenses, of radius $R$ and focal length $g$, sharing a focal plane. A 
diaphragm is placed at this plane, perforated with a small hole, of diameter 
$h$, at the point where the principal axis of the lenses
intercepts the diaphragm. 

Provided $h/g \ll \lambda/s$, where $s$ is the interferometer slit-seperation, 
the permitted deviation from horizontality of the
rays will not affect the fringe pattern observed on the interferometer screen.
Bob's interferometer is located at distance $d$. 

According to the scheme of Figure \ref{rayent}, the general nonlocal 
multi-mode vacuum state of the photons in the experiment is given by:
\begin{equation}
\label{spdc}
|\Psi\rangle = |{\rm vac}\rangle + \epsilon
(|s_{po}i_{po}\rangle + |s_{p-}i_{p+}\rangle + 
|s_{qo}i_{qo}\rangle + |s_{q-}i_{q+}\rangle   
\end{equation}
where $|{\rm vac}\rangle$ is the vacuum ground state;
$s_{po}$ and $s_{p-}$ are the photon modes on the 
$-p_{\rm side}$ and $-p_{\rm down}$ signal (Alice's) beams, emanating 
from point $p$ in the source, as shown in Figure \ref{rayent},
and $i_{po}$ and $i_{p+}$ are the photon modes on the
$p_{\rm side}$ and $p_{\rm up}$ idler (Bob's) beams, emanating from 
the same point in the source. Analogously, for the modes originating from
point $q$ on the source,  
$s_{qo}$ and $s_{q-}$ are the modes on the 
$-q_{\rm side}$ and $-q_{\rm down}$ signal beams, 
and $i_{qo}$ and $i_{q+}$ are the modes on Bob's 
$q_{\rm side}$ and $q_{\rm up}$ idler beams. The quantity
$\epsilon (\ll 1)$ depends on the
pump laser and nonlinearity in the downconverting crystal \cite{str95}.
 
Because of the narrowness of the hole,
a ray entering it diffracts to enter both slits $u$ and $v$ in the
interferometer. In the Schr\"odinger picture, let the 
diffracting wavefunction, for some ray $|X\rangle$, be: 
\begin{equation}
\label{tform}
|X\rangle \longrightarrow \alpha\left[\sin(\phi /2) |u\rangle + 
\cos(\phi /2) |v\rangle\right] + \cdot\cdot\cdot
\end{equation}
where the $\cdot\cdot\cdot$ indicate other points on the double-slit diaphragm 
where the photon could fall, but which do not concern us since they do not
pass through the double slit. 
Here $\alpha$ $( < 1 )$ depends on $h$ and the cross-section of the 
slits, $\phi$ is (some function of) $|X\rangle$'s angle of
incidence on the diaphragm as measured from the $+y$-axis whose origin is at
the hole. Note that for a ray perpendicular to the diaphragm, the amplitude 
for entering both slits is equal. Given the finite width of the 
downconverted beam,
all of which Bob's first lens is assumed to intercept, 
$\phi$ will take values between some some $\phi_0$ ($> 0$) and $180^{\circ} - 
\phi_0$. 

By virtue of the direction filter, only the rays $p_{\rm side}$ and 
$q_{\rm side}$ can fall on Bob's detector. As a result, the
 (positive mode) electric field $E^{(+)}_B$ at a point $x$ on Bob's 
detector has contributions from two of the modes: $i_{p0}$ 
falling on the hole at some angle $\phi$, and $i_{q0}$ at angle $180 - \phi$. 
The phase for each ray depends on the distance along its path. 
\begin{eqnarray}
\label{Eb}
E^{(+)}_B &=& 
\alpha\hat{i}_{po}\left( \sin(\phi /2) e^{ik(d^{\prime} + \overline{ux})}
+ \cos(\phi /2) e^{ik(d^{\prime} + \overline{vx})}\right) \nonumber \\
&+& \alpha\hat{i}_{qo}\left( \cos(\phi /2) e^{ik(d^{\prime} + \overline{ux})}
+ \sin(\phi /2) e^{ik(d^{\prime} + \overline{vx})}\right), 
\end{eqnarray}
where the "hatted" quantities represent corresponding annihilation operators,
$d^{\prime} \equiv 2g + \overline{KN} + \overline{Nv}
= 2g + \overline{LM} + \overline{Mu}$, and the overline quantities 
are distances between the named points. The four terms in right hand side of
Eq. (\ref{Eb}) can be
understood as follows. The first term is the product of three amplitudes:
for beam $p_{\rm side}$'s movement (a) from point $p$ to the hole, (b) 
from the hole to slit $u$, (c) from slit $u$ to point $x$ on Bob's screen.
Similarly with the other terms.

\subsection{Alice measures position}

Alice positions her detector on the image plane. A detection at point $y$ 
(Figure \ref{rayent}) means 
that the rays $-p_{\rm side}$ and $-p_{\rm down}$ were chosen on the signal 
beam. Alice knows that her photon originated at point $p$,
but its momentum remains indeterminate between directions $-p_{\rm side}$ and 
$-p_{\rm down}$. Correspondingly, the idler photon is left in the correlated
ray states $p_{\rm side}$ and $p_{\rm up}$. 

The (positive mode) electric field, $E^{(+)}_A$, of Alice's detector has a
contribution from both signal rays, $-p_{\rm side}$ and $-p_{\rm down}$,
that converge to $y$. Therefore, Alice's detector is given by the field:
\begin{equation}
E^{(+)}_A = \hat{s}_{po}e^{ik(2f + \overline{ry})} + 
              \hat{s}_{p-}e^{ik(\overline{py})}.
\end{equation}
The correlation function for Alice finding her photon at $y$ and Bob his at $x$ 
is $\langle E^{(+)}_A E^{(+)}_B \rangle$, where 
$\langle \cdot\cdot\cdot \rangle$ indicates
an averaging over the state vector $|\Psi\rangle$ of Eq. (\ref{spdc}).
\begin{equation}
\langle E^{(+)}_AE^{(+)}_B\rangle = \epsilon\alpha (\sin(\phi /2)
 e^{ik(2f + \overline{ry} + d^{\prime} + \overline{ux})} +
 \cos(\phi /2) e^{ik(2f + \overline{ry} + d^{\prime} + \overline{vx})})        
\end{equation}
The probability 
$P_{AB}$ of coincident detections between these detectors is given by
$|\langle E^{(+)}_A E^{(+)}_B \rangle|^2$, which is:
\begin{equation}
\label{p_visib}
P_{AB} = \epsilon^2\alpha^2
[1 + \sin (\phi) \cos (\overline{ux} - \overline{vx})].
\end{equation}
In Eq. (\ref{p_visib}), the $\sin (\phi)$ term will be 
different for different localizations of Bob's photon.

Bob's observed intensity pattern will therefore be an averaged pattern over
the closed interval $\phi \in [\phi_0, 180^{\circ} - \phi_0]$. 
Integrating over $\phi$ and assuming for
simplicity a flat profile for the converging beam 
distributed in this range (which is obtained for a laser profile that goes as
$g^{-1}\sin^2\phi$), the ensemble intensity pattern that he finds on his screen
is:
\begin{equation}
\label{I_p_visib}
I_p = I_0 \epsilon^2\alpha^2 (1 + \cos (\phi_0))(\cos (\overline{ux} - 
\overline{vx}),
\end{equation}
where $I_0$ is the idler beam intensity.
We note that the interference pattern is observed in the single 
(as against coincidence) count intensity. The reason is that $P_{AB}$ in
Eq. (\ref{p_visib}) does not depend on any Alice variables, whose presence
would have, upon being marginalized, resulted in washing out the interference
pattern. This is made possible by Bob's filter.

\subsection{Alice measures momentum}

Alice positions her detector on the focal plane. A detection at some point 
means that the signal beam is left with only parallel rays converging to 
this point. To obtain coincidence events, we need consider only signal rays 
converging to point $m$ in Figure \ref{rayent}, since Bob's filter 
permits only rays entangled to them. A detection here implies   
that the rays $-p_{\rm side}$ and $-q_{\rm side}$ were chosen on the signal 
beam.  Correspondingly, the idler photon is left in ray 
states corresponding to ray $p_{\rm side}$ and $q_{\rm side}$. Thus, Bob's
photon has a definite (horizontal) momentum, but its point of origin is
indeterminate between $p$ and $q$. 

Here Alice's detector's electric field is given by:
\begin{equation}
E^{(+)}_A = \hat{s}_{po}e^{ik(2f + \overline{rm})} + 
              \hat{s}_{qo}e^{ik(2f + \overline{tm})}
\end{equation}
Noting that $\overline{rm} = \overline{tm}$, we find:
$P_{AB} = |\langle E^{(+)}_A E^{(+)}_B \rangle |^2$ using Eqs. (\ref{spdc})
and (\ref{Eb}):
\begin{equation}
P_{AB} = 2\epsilon^2\alpha^2\left(1 + \sin\phi \right)
[1 + \cos (\overline{ux} - \overline{vx})].
\label{interf}
\end{equation}
Here, too, $P_{AB}$
depends only on the path difference, $\overline{ux} - \overline{vx}$, from the
slits to $x$. It does not depend on any of Alice's variables. 

The intensity pattern observed by Bob in the single counts is therefore
given by the function:
\begin{equation}
\label{I_m_visib}
I_m = 2I_0\epsilon^2\alpha^2 (1 + \cos\phi_0)
[1 + \cos (\overline{ux} - \overline{vx})].
\end{equation}
This has a visibility function of 1.0 in contrast to $\cos\phi_0$, found in 
the case of $I_p$ given in Eq. (\ref{I_p_visib}). Furthermore, $I_m$ is about
twice more intense than $I_p$. Therefore the intensity and visibility of Bob's 
single count interference pattern 
are affected by what Alice observes. She transmits one classical bit of
information nonlocally. 

The experiment is essentially a delayed choice experiment \cite{wheeler}, 
as applied to entangled particles instead of a single particle. Alice forces
Bob's entangled particle to behave like a wave or particle by measuring a 
wave (i.e, momentum) or particle (i.e, path) property of her photon. 
The interesting part is that she may delay her choice of which aspect
to manifest until after the photons have left the source. 

A more dramatic demonstration of the signaling is obtained by minimizing
diffraction at the hole by increasing size $h$. Position
measurement by Alice will result in an idler ray passing through 
only one of the two slits.
No interference will result. On the other hand, her momentum measurement will
result in an interference because the optics will ensure that Bob's photon
passes through both slits
thereby producing an interference. Thus, Bob simply checks for the presence or
absence interference pattern to infer Alice's action. One complication here is
that a larger hole size would allow non-horizontal momenta to enter the
interferometer, thereby reducing the visibility. Therefore the hole cannot be 
too large. The condition
$1 \ll h/\lambda \ll g/s$ ensures that these criteria are satisfied. But it 
could require larger $g$ than may be feasible for entangled beams of finite
coherence length. Nevertheless, the origin of the signaling can be more
simply understood for this case, as done in the next subsection.

\section{Understanding the classical signal}

The no-signaling theorem implies that statistically the outcomes for Bob's 
particle is independent of Alice's action \cite{nosig}. It is of interest
to know how the above experiment circumvents it. The answer is basically
that the proofs of no-signaling assume that both Alice and Bob make complete 
measurements, whereas in the above experiment, their interferometric 
measurement is incomplete, because a detection does not uniquely indicate an
eigenmode. Furthermore, Bob employs a filter, which restricts
his observation to the incomplete measurement projector $\hat{M}_B =
|i_{p0}\rangle\langle i_{p0}| + |i_{q0}\rangle\langle i_{q0}|$. Another
factor is a subtlety concerning the scope of the complementarity principle in
multi-particle interferences.

In a position measurement, Alice collapses $|\Psi\rangle$ with 
one of the incomplete measurement projectors $\hat{P}_1 =
|s_{p0}\rangle\langle s_{p0}| + |s_{p-}\rangle\langle s_{p-}|$ and $\hat{P}_2 =
|s_{q0}\rangle\langle s_{q0}| + |s_{q-}\rangle\langle s_{q-}|$.
In a momentum measurement, her operators are $\hat{M}_1 =
|s_{p0}\rangle\langle s_{p0}| + |s_{q0}\rangle\langle s_{q0}|$ or $\hat{M}_2 =
|s_{p-}\rangle\langle s_{p-}| + |s_{q-}\rangle\langle s_{q-}|$. If Alice
measures position, the state of photons observed by Bob is 
$\hat{M}_B\hat{P}_i|\Psi\rangle$. This is given by the statistical mixture:
\begin{equation}
\label{statmixp}
\rho_p = \frac{1}{2}\left(|s_{p0}i_{p0}\rangle\langle s_{p0}i_{p0}| +
|s_{q0}i_{q0}\rangle\langle s_{q0}i_{q0}|\right).
\end{equation}
As there are no cross-terms between the paths,
Bob observes no interference in this case. 

On the other hand, if Alice
measures momentum, the state of photons observed by Bob is:
\begin{equation}
\label{state}
\hat{M}_B\hat{M}_1|\Psi\rangle = 
(1/\sqrt{2})(|s_{p0}i_{p0}\rangle + |s_{q0}i_{q0}\rangle ),
\end{equation}
since $\hat{M}_B\hat{M}_2|\Psi\rangle = 0$.
A direct application of complementarity would suggest that Bob's rays 
$|i_{p0}\rangle$ and $|i_{q0}\rangle$ in Eq. (\ref{state})
cannot produce an interference pattern
because their entanglement with the signal photon,
whose states $|s_{p0}\rangle$ and $|s_{q0}\rangle$
are mutually orthogonal, makes them distinguishable. This is equivalent to 
saying that interference is not possible because tracing 
over the signal states \cite{can78} in the density matrix 
$\hat{M}_B\hat{M}_1|\Psi\rangle\langle \Psi |\hat{M}_1^{\dag}\hat{M}_B^{\dag}$
results in $\rho_p$. {\em But this conclusion is not
supported by the experimentally attested two-particle coincidence 
interferences} \cite{ghosh,str95,zei00}.
The more rigorous approach would be to analyze
interference in terms of the phase accumulated by Bob's particle's wavefunction
along each path, without reference to external states \cite{ste90}.
Though complementarity is an excellent thumb-rule to elucidate many 
non-classical effects, it is essentially
a qualitative idea, and its use warrents some caution \cite{srik_complem}.

The observation of two-particle interference implies that the phase 
contribution to the wavefunction on a path is accounted for by the distance 
traversed by a ray along its path. No phase (or uniform phase) is picked up
at the detector. Bob's filter-interferometer system is crucial as it ensures 
that only two horizontal modes are allowed (i.e, it implements $\hat{M}_B$). 
This fixes the point of Alice's detection in coincidences.
Therefore, the relative phase 
in Bob's wavefunction at the two slits remains constant, 
permitting the formation of an observable interference pattern on his screen. 
For the set-up in Figure \ref{rayent}, the relative phase vanishes.
Classical signaling is the direct consequence of the distinguishability of
$\hat{M}_B\hat{M}_1|\Psi\rangle$ in Eq. (\ref{state}) from $\rho_p$ in Eq.
(\ref{statmixp}). We note that the single mode probabilities for states
$|i_{p0}\rangle$ and $|i_{q0}\rangle$ are the same for both cases. Thus, 
the no-signaling theorem, within the scope of its implicit assumption, is
not violated. Also, for the same reason, no violation of probability 
conservation occurs.

\section{Discussion}\label{discuss}

The above experiment aims to prove a much stronger condition 
about nonlocal correlations than does Bell's theorem, in two ways. First:
unlike Bell's theorem, it does
not assume the reality of underlying variables; just the tested principles
of quantum mechanics and quantum electrodynamics suffice. 
Second: the nonlocal influence is shown 
to transmit classical information, which means that the correlations are
not uncontrollable. Of course, this leads to the problematic situation that 
the effective speed $v_{\rm eff}$ of the transmission of this nonlocal 
classical signal can be made arbitrarily large. In one sense this need not
be surprising: for quantum mechanics is a non-relativistic theory. 
The no-signaling theorem is based on quantum mechanical unitarity and
assumptions of quantum measurement rather than
relativisitc signal locality. If the
instant of Alice's choice is $t_a$ and that of Bob's measurement is
$t_b$, where $t_b \ge t_a$, then
$v_{\rm eff} = (d + 4f)/(t_b - t_a)$, 
where the upper 
limit occurs in the situation where Alice delays her choice until just before 
her photon reaches her. Since this can be made arbitrarily large by increasing 
$4f$ and $d$ and/or decreasing the time difference, suggesting that 
nonlocality does not prohibit a superluminal classical signal,
it is of interest to examine factors inhibiting such a possibility.

The situation is not improved by considering a no-collapse scenario like
the relative-state (or many-worlds) interpretation \cite{eve50}, because each
branching universe would have to contend with the classical signal. 
Requiring collapse to be subluminal (even simply non-instantaneous)
would imply significant changes to quantum theory, and
furthermore permit non-conservation of entangled quantities. By 
demonstrating a classical transfer of information, the above experiments
would corraborate the objective nature of state vector reduction.

It appears that to 
restore causality we would need the light to somehow decohere into 
disentangled momentum pointer states \cite{zeh70}
before reaching Bob's double-slit so that he will always find a Young's
double-slit interference pattern irrespective of Alice's action. But it is not 
clear how such a decoherence can be brought about. Perhaps somehow
spacetime itself 
would act as a measuring environment to the system, thereby decohering it, in 
order to safeguard its own causal structure? Such an explanation is related to 
an incompleteness in quantum
mechanics as a formal axiomatic theory \cite{srik}, and would probably require
new axioms in the theory. But it would also reveal an unexpected
connection between quantum information and decoherence. In any case, the 
technical feasibility of the above described thought experiment permits a 
facile test for nonlocal communication of the type claimed in this paper.

\section{Conclusion}

The question raised in the title of this article is answered in the
affirmative. It is unclear how this is to be reconciled with signal locality.
Perhaps this points to the need for new physics to understand state vector
"collapse", more so in nonlocal cases. 

\acknowledgements

I am thankful to Dr. R. Tumulka, Dr. R. Plaga, Dr. M. Steiner, 
Dr. J. Finkelstein and Dr. C. S. Unnikrishnan for their constructive criticism.
I thank Ms. Regina Jorgenson for her valuable suggestions.

\newpage
 
\end{document}